\documentclass[
amsmath,amssymb,
prd,nofootinbib,floatfix,12pt,
]{revtex4}
\usepackage{amsmath,amssymb,amsfonts}
\usepackage[dvips]{color}
\usepackage{xcolor}
\usepackage{dsfont}
\usepackage{mathrsfs}
\usepackage{float}
\usepackage{hyperref}
\usepackage{graphicx}
\usepackage{graphics}
\usepackage{setspace}
\usepackage{lmodern}
\usepackage{tikz}
\usepackage{cancel}
\usepackage{placeins}
\hypersetup{colorlinks,citecolor=blue,urlcolor=blue,linkcolor=black}
\newcommand\beq{\begin{eqnarray}}
\newcommand\eeq{\end{eqnarray}}

\allowdisplaybreaks
\begin{document}
\renewcommand{\theequation}{\arabic{section}.\arabic{equation}}
\renewcommand{\thefigure}{\arabic{section}.\arabic{figure}}
\renewcommand{\thetable}{\arabic{section}.\arabic{table}}

\title{\large
A sensitivity target for an impactful\\ Higgs boson self coupling measurement}

\author{Prudhvi N.~Bhattiprolu and James D.~Wells}

\affiliation{\it
  {Leinweber Center for Theoretical Physics,\\ University of Michigan, Ann Arbor, MI 48109, USA}
}

\begin{abstract}\normalsize 
We argue that a measurement of the Higgs boson self-coupling becomes particularly meaningful in a large and important class of theories when its sensitivity is within 40\% of its Standard Model value.  This constitutes a target for a future impactful experimental achievement. It is derived from recently obtained results of how extreme the differences can be between effective field theory operator coefficients when their origins are from reasonable custodial-violating theories beyond the Standard Model. 
\end{abstract}

\maketitle


\setcounter{footnote}{1}
\setcounter{figure}{0}
\setcounter{table}{0}

\section{Introduction\label{sec:introduction}}
\setcounter{equation}{0}
\setcounter{figure}{0}
\setcounter{table}{0}
\setcounter{footnote}{1}

The precise measurement of the Higgs self-coupling is essential for understanding the electroweak symmetry breaking mechanism and the stability of the Higgs potential. Deviations from the Standard Model (SM) predictions for the Higgs self-coupling can signal the presence of new physics beyond the SM (BSM), providing crucial insights into the fundamental nature of our universe. The parameter $\kappa_\lambda \equiv c_{h^3}/c_{h^3}^\text{SM}$, defined as the ratio of the Higgs self-coupling to its SM value, is currently constrained to be $-0.4 < \kappa_\lambda < 6.3$ at 95\% CL with the assumption that only the Higgs self-coupling is modified by new physics \cite{ATLAS:2022jtk}.

In this work, we set a search target for the Higgs self-coupling in BSM scenarios where the ratio of fractional deviations in the Higgs self-coupling $\delta_{h^3} \equiv \Delta c_{h^3}/c_{h^3}^\text{SM}$ to those in the single-Higgs coupling, such as Higgs to vector bosons $\delta_{VV} \equiv \Delta c_{VV}/c_{VV}^\text{SM}$, can be large. 
Of course, any improvement on the measurement of the Higgs self-coupling is important, as it would explore the parameter space of BSM scenarios known and unknown. Much work has been done in the literature to make these connections between the many theories and the Higgs self coupling measurement expectation~\cite{Durieux:2022hbu,Barger:2003rs, Grojean:2004xa, Gupta:2013zza, DiVita:2017eyz, Giudice:2007fh, Liu:2016idz, Hollik:2001px,DiLuzio:2017tfn,DiLuzio:2016sur, Chang:2019vez, Kanemura:2002vm, Carvalho:2017vnu, Falkowski:2019tft, Agrawal:2019bpm, Abu-Ajamieh:2020yqi, Kribs:2017znd, deBlas:2014mba, Degrassi:2017ucl,Buttazzo:2015bka,Ginzburg:2015yva, Dawson:2015oha, Azatov:2015oxa, Goertz:2015dba}.

In this article we would like to go a step further and use known results to construct a target for measurement of the Higgs self coupling.
The basic insight is to realize that it is unexpected that the effective theory that results from integrating out a ultraviolet (UV) theory should have only one operator that affects only one observable. In other words, one does not expect that the coefficient of the effective theory operator $|H|^6$ that affects most directly the Higgs boson self-coupling $\delta_{h^3}$ should affect no other operator. In fact, when the effective theory arises from a reasonable UV theory, the $|H|^6$ coefficient cannot be too extremely different than at least some other operator coefficients that give rise to $\delta_{VV}$ deviations. 

In this spirit, reference~\cite{Durieux:2022hbu} derived an upper bound on $\xi \equiv \left| \delta_{h^3}/\delta_{VV}\right|$ in generic UV completions of the SM:
\beq
\xi \equiv \left| \delta_{h^3}/\delta_{VV}\right| \>\ \lesssim \>\ 600.
\eeq
We are agnostic to specific UV completions that might give rise to extreme values near 600, but merely utilize this as a reasonable upper limit. We will show below that this maximum value of $\xi$, combined with precision electroweak measurements already taken, implies that the Higgs self coupling should be within 40\% of the Standard Model value. This is our claimed necessary target sensitivity to have any hope of finding a deviation beyond the Standard Model in the Higgs self coupling.

\section{EFT characterization of the deviations in the Higgs self-coupling\label{sec:h3EFT}}
\setcounter{equation}{0}
\setcounter{figure}{0}
\setcounter{table}{0}
\setcounter{footnote}{1}

In our analysis, we characterize the deviations in the Higgs self-coupling and the single-Higgs couplings in terms of dimension-6 operators constructed solely from the $SU(2)_L$-doublet Higgs field $H \equiv \left(0 \ \ (v + h)/\sqrt{2}\right)^T$ and its covariant derivative $D_\mu H$. Here, $v \equiv 1/\sqrt{\sqrt{2} G_F} \simeq 246$ GeV is the Higgs vacuum expectation value (VEV) that is experimentally determined from the Fermi coupling constant $G_F$.

Specifically, we restrict our analysis to the following two most important operators:
\beq
\mathcal{L} &\supset& -\frac{1}{\Lambda_6^2} |H|^6 - \frac{2 \alpha T}{v^2} |H^\dagger D_\mu H|^2,
\label{eq:H6+T}
\eeq
where $\alpha$ is the electromagnetic fine-structure constant, $\Lambda_6$ is the cutoff scale associated with the $|H|^6$ operator, and $T$ is one of the Peskin-Takeuchi parameters \cite{Peskin:1991sw}.
The $|H|^6$ operator is unique as it only affects the Higgs self-coupling without generating other effective operators through renormalization group evolution \cite{Jenkins:2013zja}.
In addition to the $T$ parameter, one can also include operators such as
$\left( \partial_\mu |H|^2 \right)^2$ and $|H|^2 |D_\mu H|^2$ (see, e.g, Refs.~\cite{Durieux:2022hbu,Wells:2015uba}). However, we focus exclusively on the $T$ parameter in our analysis because it is one of the most precisely measured parameters in precision electroweak data, providing stringent constraints on the deviations in the Higgs self-coupling.

The $|H|^6$ term modifies the scalar potential as follows:
\beq
\label{eq:potential}
V(H) &=& m_H^2 |H|^2 + \lambda |H|^4 + \frac{1}{\Lambda_6^2} |H|^6,
\eeq
that has a minimum at
\beq
\langle |H|^2 \rangle \equiv \frac{v^2}{2} &=& \frac{\lambda \Lambda_6^2}{3} \left( \sqrt{1 + \frac{3 {\hat v}^2}{\lambda \Lambda_6^2}} - 1 \right),
\eeq
with $m_H^2 < 0$, $\lambda > 0$, and ${\hat v}^2 \equiv -m_H^2/\lambda$ that minimizes $V^\text{SM}(H)$ (i.e., the first two terms in eq.~(\ref{eq:potential})). This solution recovers the SM in the large $\Lambda_6$ limit \cite{Barger:2003rs,Grojean:2004xa}. Therefore, the $|H|^6$ term induces a small shift in the Higgs VEV $v$ that minimizes the full potential:
\beq
{\hat v}^2 &=& v^2 \left( 1 + \frac{3 v^2}{4 \lambda \Lambda_6^2} \right),
\eeq
and also modifies the physical Higgs boson mass and self-interaction couplings.

On the other hand, the $T$ parameter leads to a non-canonical kinetic term for the physical Higgs boson $h$:
\beq
\mathcal{L} &\supset& \frac{1}{2 \mathcal{N}^2} (\partial_\mu h)^2,
\eeq
where $\mathcal{N} = 1/\sqrt{1 - \alpha T}$, along with modifications to the $Z$ boson mass and the couplings of Higgs boson to the $Z$ boson and itself.
To restore the canonical form of the kinetic term for $h$, we can redefine $h \rightarrow \mathcal{N} h$, which further modifies all the interactions of the physical Higgs boson.

In the presence of both operators in Eq.~(\ref{eq:H6+T}), the Higgs mass and the Z boson mass receive contributions:
\beq
m_h^2 &=& 2 \mathcal{N}^2 \lambda v^2 \left(1 + \frac{3 v^2}{2 \lambda \Lambda_6^2}\right),\\
m_Z^2 &=& \frac{g^2 v^2}{4 c_W^2 \mathcal{N}^2},
\eeq
where $g$ is the $SU(2)_L$ gauge coupling, and $c_W$ is the cosine of the weak mixing angle. Canonically normalizing the Higgs kinetic energy, fixing $m_h$ and $m_Z$ to their corresponding pole masses, and expanding the potential in large $\Lambda_6$ and small $\alpha T$, the effective Lagrangian now contains:
\beq
\mathcal{L} &\supset&
-\frac{m_h^2}{2 v} \left[\left(1 + \frac{2 v^4}{m_h^2 \Lambda_6^2} + \frac{\alpha T}{2}\right) h^3 + \frac{2 \alpha T}{m_h^2} h (\partial_\mu h)^2\right]\notag\\
&&+ \frac{m_Z^2}{v} \left(1 - \frac{\alpha T}{2} \right) h Z_\mu Z^\mu.
\eeq
There are also contributions to, for example, the couplings of $h$ to fermions and $W$ bosons from the $T$ parameter, which we do not show here. Here we take the $hZZ$ coupling as a representative for the Higgs to vector boson coupling.

Therefore, the fractional deviations in the Higgs self-coupling ($h^3$) and the Higgs to vector boson coupling ($hZZ$) are
\beq
\delta_{h^3} &=& \frac{2 v^4}{m_h^2 \Lambda_6^2} + \frac{\alpha T}{2},\\
\delta_{VV} &=& - \frac{\alpha T}{2},
\eeq
respectively.  In the kappa framework, $\kappa_\lambda = 1 + \delta_{h^3}$.


\begin{figure}
 \begin{center}
    \includegraphics[width=0.9\columnwidth]{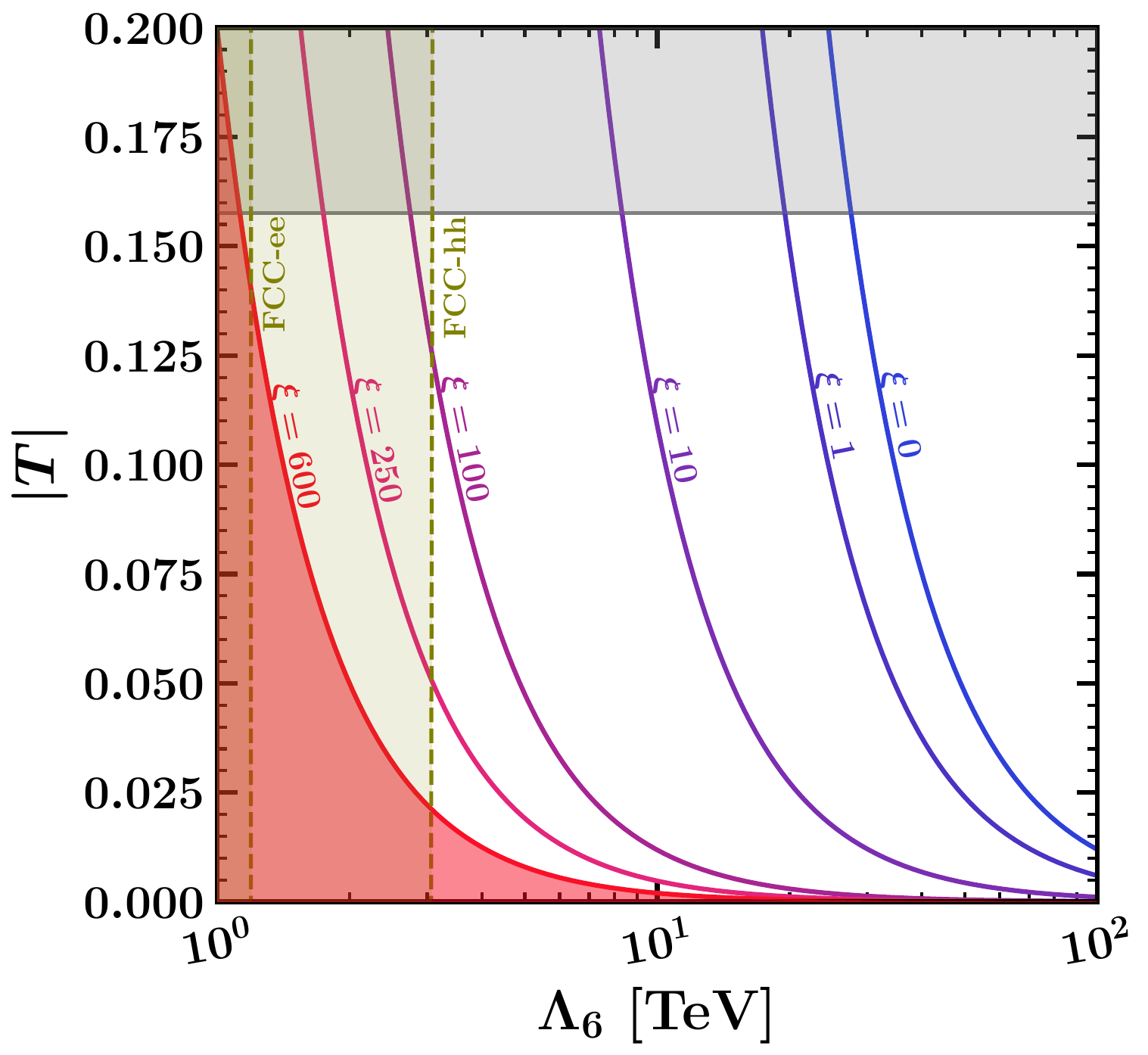}
 \end{center}
 \caption{The $T$ parameter as a function of the $\Lambda_6$ scale for various values of $\xi \equiv \left|\delta_{h^3}/\delta_{VV}\right|$. The gray shaded region corresponds to the current 95\% CL bound on the $T$ parameter \cite{ParticleDataGroup:2020ssz}, and the red shaded region corresponds to the theoretical upper bound on $\xi$ \cite{Durieux:2022hbu}. The olive shaded regions correspond to the $1\sigma$ sensitivity on the Higgs self-coupling at various future colliders \cite{DiMicco:2019ngk} as labeled.
\label{fig:Tparameter}}
\end{figure}


\begin{figure*}[t!]
   \begin{minipage}[]{0.7\columnwidth}
    \centering
    \includegraphics[width=\columnwidth]{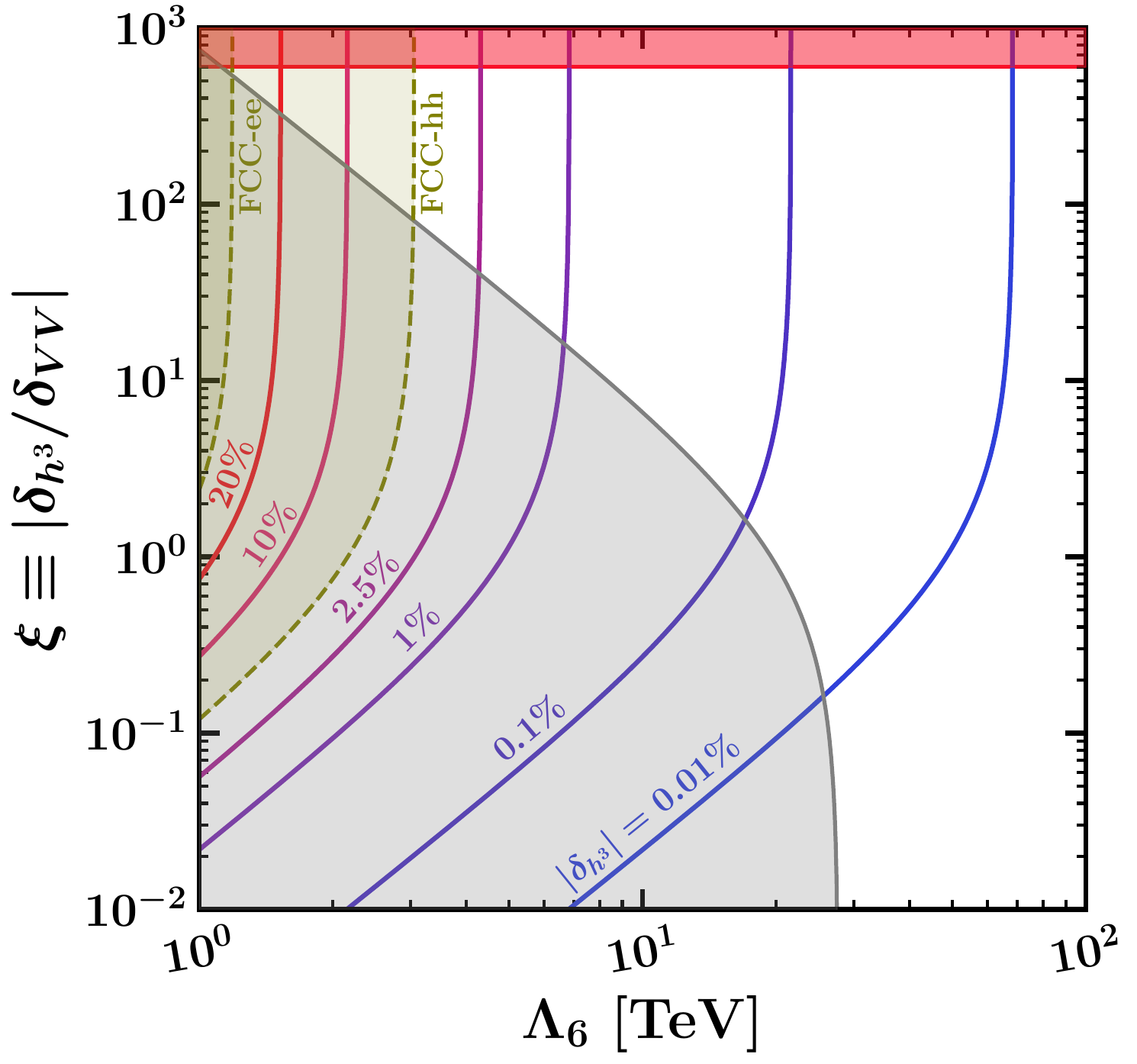}
  \end{minipage}
  \begin{minipage}[]{0.7\columnwidth}
    \centering
    \includegraphics[width=\columnwidth]{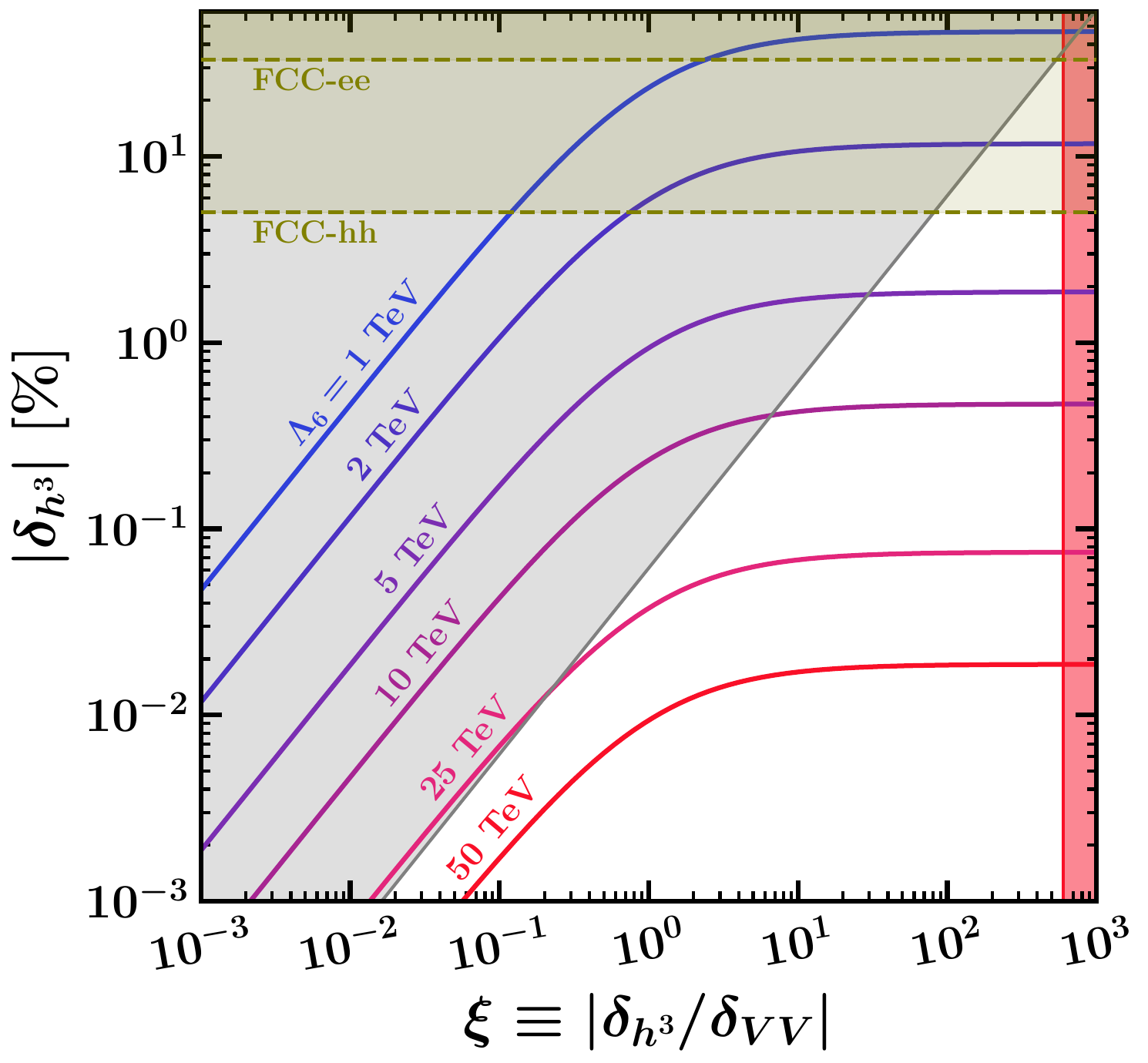}
  \end{minipage}
 \caption{Top: $\xi$ as a function of $\Lambda_6$ for various values of $|\delta_{h^3}|$. Bottom: $|\delta_{h^3}|$ as a function of $\xi$ for various values of $\Lambda_6$. In both panels, the gray shaded region corresponds to the current 95\% CL bound on the $T$ parameter \cite{ParticleDataGroup:2020ssz}, and the red shaded region corresponds to the theoretical upper bound on $\xi$ \cite{Durieux:2022hbu}. And, the olive shaded regions correspond to the $1\sigma$ sensitivity on the Higgs self-coupling at various future colliders \cite{DiMicco:2019ngk} as labeled.
\label{fig:deltah3}}
\end{figure*}

\section{BSM targets for the Higgs self-coupling\label{sec:BSMtargets}}
\setcounter{equation}{0}
\setcounter{figure}{0}
\setcounter{table}{0}
\setcounter{footnote}{1}

Legitimate BSM theories can produce large values of $\xi \lesssim 600$ \cite{Durieux:2022hbu}.
By focusing on deviations characterized by $|H|^6$ and $T$ operators, we can identify specific BSM targets for the Higgs self-coupling.

In our analysis, we use precision electroweak data, particularly the highly constrained $T$ parameter, 
to constrain the parameter space.
In terms of $\xi = \left|\delta_{h^3}/\delta_{ZZ}\right|$, the $T$ parameter and the fractional deviation in the Higgs self-coupling can be expressed as:
\beq
|T| &=& \frac{1}{1 + \xi} \ \frac{4 v^4}{\alpha m_h^2 \Lambda_6^2},\\
|\delta_{h^3}| &=& \frac{\xi}{1 + \xi} \ \frac{2 v^4}{m_h^2 \Lambda_6^2},
\eeq
respectively.
The $T$ parameter is currently constrained at 95\% CL to be $|T| \lesssim 0.16$ \cite{ParticleDataGroup:2020ssz}. This, together with the requirement that $\xi \lesssim 600$, implies that
\beq
\Lambda_6 \>&\gtrsim&\> 1.1 \text{ TeV},\\
|\delta_{h^3}| \>&\lesssim&\> 40\%.
\eeq
Any improvement in the measurement of the $T$ parameter, such as from a $Z$-pole run at the FCC-ee \cite{deBlas:2019rxi}, would lower the above target for $|\delta_{h^3}|$ by a factor of $(|T|/0.16)$.
Note that this target number for $|\delta_{h^3}|$ could be higher and fluid in the context of a class of BSM theories with strong custodial symmetry preservation \cite{Durieux:2022hbu}--in that case, any improvement in the self-coupling measurement would be meaningful.
In comparison, future colliders such as the FCC-ee and FCC-hh would be able to measure $|\delta_{h^3}|$ to levels of about 33\% and 5\%, respectively \cite{DiMicco:2019ngk}. The FCC-ee would be a marginally impactful measurement according to our analysis, and the FCC-hh would be a highly impactful measurement.

In Figure~\ref{fig:Tparameter}, we show the $|T|$ parameter as a function of the $\Lambda_6$ scale for various $\xi$ values.
Figure~\ref{fig:deltah3} consists of two panels.
The top panel shows various contours of $|\delta_{h^3}|$ in the $\Lambda_6$ vs $\xi$ plane, and the bottom panel shows various contours of $\Lambda_6$ in the $\xi$ vs $|\delta_{h^3}|$ plane.
In both figures, the gray shaded regions correspond to the constraint from the $T$ parameter \cite{ParticleDataGroup:2020ssz}. And, the red shaded regions correspond to the theoretical upper bound on the ratio of fractional deviations in the Higgs self-coupling to those in the Higgs to vector boson coupling, $\xi$, in generic UV completions \cite{Durieux:2022hbu}.
The olive shaded regions show the projected sensitivity on the Higgs self-coupling at FCC-ee and FCC-hh \cite{DiMicco:2019ngk}.

As evident from Figures~\ref{fig:Tparameter} and \ref{fig:deltah3}, the $T$ parameter
is highly constraining. Nonetheless, future colliders will begin to probe the parameter space of BSM theories with large $\xi$ values. Specifically, the FCC-ee and FCC-hh will be able to explore scenarios with $\xi \gtrsim 535$ and $\xi \gtrsim 80$, respectively, and $\Lambda_6 \lesssim 1.2$ TeV and $\Lambda_6 \lesssim 3$ TeV, respectively.

\section{Conclusion\label{sec:conclusion}}
\setcounter{equation}{0}
\setcounter{figure}{0}
\setcounter{table}{0}
\setcounter{footnote}{1}

In this study, we analyzed the relationship between operators of an effective theory that impact the relatively unconstrained Higgs self coupling and those that impact the highly constrained precision electroweak data in order to understand what the maximum experimentally allowed deviation is in the Higgs boson self coupling compared to its Standard Model value. Our analysis found a target of $40\%$ sensitivity for the Higgs self-coupling determination with respect to the Standard Model value, particularly in reasonable custodial-violating theories, in order to have viable expectation of possible deviations seen. This maximum deviation is derived assuming a maximum ratio of the Higgs self-coupling relative to the single-Higgs couplings, $\xi \equiv \left|\delta_{h^3}/\delta_{VV}\right| \lesssim \>\ 600$, allowed by generic UV completions of the Standard Model.

\medskip
{\it Acknowledgments:}
We thank Zhengkang (Kevin) Zhang for helpful discussions.
This research was supported in part through computational resources and services provided by
Advanced Research Computing (ARC), a division of Information and Technology Services (ITS) at
the University of Michigan, Ann Arbor. This work is supported by the Department of Energy under
grant number DE-SC0007859.

\bibliographystyle{plain}
\bibliography{ref}

\end{document}